\documentclass[12pt]{article}
\usepackage{amsfonts,amssymb,graphicx}
\newcommand{\binomial}[2]{\left(\begin{array}{c}#1\\#2\end{array}\right)}
\newcommand{\w}{\omega}
\newcommand{\W}{\Omega_s}
\newcommand{\legendre}[2]{\left(\frac{#1}{#2}\right)}
\def\R{I\!\!R}

\def\Z{I\!\!\!\!Z}

\def\F{{\cal{F}}}

\def\E{{\rm E}\,}
\def\P{{\rm P}\,}

\def\O{\Omega}
\def\o{\omega}
\def\t{\theta}

\def\S{\Sigma}
\def\s{\sigma}

\def\vsni{\vskip 0.2cm}

\def\be{\begin{equation}}
\def\ee{\end{equation}}

\def\mod{\hbox{\rm mod\ }}

\def\\{\hfill\break}
\def\={{ \; \equiv \; }}

\def\t_N{\tilde{\Z}_N}

\begin{document}

\title{Statistics of energy levels and zero temperature dynamics for
deterministic spin models with glassy behaviour}

\author{M.Degli Esposti$^{a)}$, C.Giardin\`a$^{b)}$ , 
S.Graffi$^{a)}$, 
S.Isola$^{c)}$
\\
{\small $^{a)}$ Dipartimento di Matematica} \\
    {\small Universit\`a di Bologna, 
    40127 Bologna, Italy}
\\
   {\small $^{b)}$ Dipartimento di Fisica}\\
    {\small Universit\`a di Bologna, 
    40126 Bologna, Italy }
\\
   {\small $^{c)}$ Dipartimento di Matematica e Fisica}\\
   {\small Universit\`a di Camerino, 
    62032 Camerino, Italy} \\
{\footnotesize E-mail: \ttfamily desposti@dm.unibo.it,
giardina@bo.infn.it,  graffi@dm.unibo.it, isola@campus.unicam.it} }
\date{}
\maketitle
\begin{abstract}\noindent
We consider the zero-temperature dynamics for the infinite-range, non
translation invariant one-dimensional spin model introduced by Marinari,
Parisi and Ritort to generate glassy behaviour out of a 
deterministic interaction. It is shown that there can be a large number
of
metatastable (i.e., one-flip stable) states with very small overlap
with the ground state but
 very close in energy to it, and that their total number increases
exponentially
with the size of the system.
\end{abstract}
\section{Introduction}
A main issue in glassy systems is the analogy
between glass-forming liquids and discontinuous spin-glasses, first
pointed out in the pioneering works by Kirkpatrick, Thirumalai
and Wolynes \cite{KTW}. 
In both cases the thermodynamical 
properties can be indeed related to the dynamical evolution in an energy 
landscape. 
In  liquid theory one can define the notion of {\em 
inherent 
structures} \cite{SW} (local minima of the potential energy,
each one surronded by its attraction basin or valley)
and {\em configurational entropy}, i.e. the logarithm of the number of
these minima divided by the number of particles in the system.
Then the low-temperature dynamical evolution can be described as 
a superposition of an intra-basin ``fast'' motion and a ``slow'' 
crossing of energy barriers. If the 
temperature of the system is small enough, namely less than the 
Mode Coupling critical temperature $T_{MC}$, the system gets 
trapped in one of the basins. Since the number of energy minima 
diverges exponentially with the size of the system, a 
thermodynamic transition can be associated with an entropy 
crisis: the Kauzmann temperature $T_{K}$ of the glassy transition 
corresponds to the vanishing of the configurational entropy.
We refer the reader to \cite{MP} 
for an overview on equilibrium 
thermodynamics of glasses.

Consider now the class of discontinuous spin glasses, i.e. the mean-field 
models  involving a random $p$-spin interaction. 
Also  these models show  a dynamical transition at a temperature 
$T_D$ (corresponding to  $T_{MC}$) where dynamical ergodicity 
breaks down;  a thermodynamic entropy-driven transition takes place at a lower 
temperature $T_{1RSB}$ (corresponding to $T_K$), at which replica 
symmetry breaks down with a ``one step'' pattern.
Here the local minima of the free energy  correspond to  
the solutions of the mean field TAP equations. Anyway at temperature
$T = 0$ metastable states with respect to any dynamics reduce 
to $1$-spin-flip stable states.

The main gap in the analogy between structural glasses and 
discontinuous 
spin-glasses is that in the latter models, unlike the former, the couplings 
between  spins are  quenched random
variables. A significant, recent step in filling this gap has been made by 
 the introduction of the {\em deterministic}, i.e.
non-random, spin models which show a dynamical phase transition with a
discontinuous order parameter and an equilibrium phase transition at a lower 
temperature
associated with the vanishing of the high-temperature entropy
\cite{{MPR},{BM},{BDGU},{NM},{PP}}. It is the high degree 
of frustation among the couplings, not the
disorder, to generate a huge number of metastable states and thus 
  the glassy behaviour. The discovery of these models proved  that
disorder is not necessary  to reproduce a complex free energy 
landscape. 

Metastables states in infinite-range disordered 
spin-glasses
have been extensively studied, both in the SK model 
 where the number of $1$-spin-flip stable 
states 
scales like $\exp (0.1992 N)$ \cite{{TE},{BrM},
{DGGO},{MPV}},  $N$ being the size of the system (number of
spins in the one-dimensional case), and  in 
$p$-spin interaction spin-glasses \cite{OF}.  

Here we deal with the same
question in deterministic models. By probabilistic arguments we will obtain,
for the models introduced  in \cite{MPR}, 
a  lower bound on the number of $1$-spin-flip stable states, 
which increases exponentially  with the size  of the system. 
Hence this deterministic model
exhibits the main feature of glassy behaviour.  

The paper is organized as follows: in Section 2 we review the basic
properties of the   model we consider, the {\em sine model} of \cite{MPR}, a
deterministic, one dimensional chain of $N$ spins with long-range oscillating
interaction.  In Section 3 we study the limiting
distribution of the rescaled  energy density, showing that it gets
$\delta$-distributed in the thermodynamic limit. This property, which holds
for the Curie-Weiss case, is an indication of the mean-field nature of the
model. In Section 4 we deal with  the state space of the
system, computing explicitly (for $N$ prime analytically, for other values of
$N$ numerically) the distribution of the energy levels by flipping one spin at
a time;  among other things we show that there can be a large number
of states with almost zero overlap with the ground state but very close
in energy to it.  Finally, in Section 5 we derive the main result of
the paper, that is a lower exponential bound for the number of metastable
states at temperature $T = 0$.  \vskip 0.2cm
\textbf{Acknowledgments:} M.~D.E. contributed to this paper during his
visit to the
School of Mathematics of the Georgia Institute of Technology, whose
support and excellent working conditions he gratefully acknowledges.

C. G. wants to thank the Department of Physics of Brown University,
where part of this work was performed, for the
kind hospitality during his visit under the exchange program between the
Bologna and Brown
Universities .
\section{Orthogonal interaction matrices and the sine model}
The basic setup is a probability space $(\S_N, \F_N,\P_N)$. 
The sample space $\S_N$ is the configuration space, i.e. 
$\S_N=\{-1,1\}^N$ whose elements are
the sequences $\s=(\s_1,\cdots ,\s_N)$ with $\s_i=\pm 1$; $\F_N$
is  the finite
algebra with $2^{2^N}$ elements, and the {\sl a priori} (or {\sl 
infinite-temperature}) probability measure $\P_N$
is given by
\be
\P_N(C)={1\over 2^N} \sum_{\s \in C} 1.
\ee
The Hamiltonian is the function on $\S_N$ defined as
\be
\label{hamiltonian}
H(\s)=-{1\over 2}\sum_{xy}J_{xy}\,\s_x\, \s_y = -{1\over 2}<J\s,\s>
\ee
where $J=(J_{xy})$ is a 
symmetric real orthogonal  $N\times N$ matrix given from 
the outset.
Although many of the results presented here will hold  for
a generic symmetric  orthogonal matrix (e.g.
of the form $J=OLO^{T}$ with $L$ a diagonal matrix whose elements are 
$\pm 1$ and $O$
a generic orthogonal matrix chosen at random w.r.t. the Haar measure on 
the orthogonal
group) in what follows we shall examine the following particular 
example known as the
{\sl sine model}:
\be
\label{sine}
J_{xy} = {2\over \sqrt{2N+1}} \sin \left({2\pi xy \over 2N+1}\right).
\ee
which satisfies (we assume $N$ odd) \footnote{One might also consider
interaction matrices  with zero diagonal terms,
recovering orthogonality in large $N$ limit. This amounts to put the 
average energy equal to zero
(instead of $-1/2$) and may be convenient for particular purposes (see 
Section \ref{limiting_distribution}).}
\be
\label{orth}
J \, J^{T} = {\rm Id}\quad \hbox{and}\quad 
\sum_{x=1}^N J_{xx}=\sum_{x=1}^N J_{xx}^2=1.
\ee
This  model has been introduced by Marinari, Parisi and
Ritort as a deterministic system with high frustration
(competiting interactions with different signs and strengths) able to 
reproduce the complex thermodynamical behaviour typical of spin glasses 
\cite{MPR}. 
It has been investigated  analytically in the high-temperature regime, 
(through 
an high-temperature expansion), and numerically also in the 
low-temperature
phase (using Montecarlo annealing). The analytical study revealed
the existence of a static phase transition at a temperature
$T_S = 0.065$ where the high-temperature entropy
vanishes, while evidence of
the existence of a higher temperature $T_D = 0.134$ where the system undergoes
a dynamical transition of second order (i.e. with a jump in the specific heat) with
a discontinuous order parameter has been put forward by numerical analysis. It has
also been shown, using the replica formalism, that most of the  thermodynamical
properties of this model are the same as those of a generic symmetric orthogonal
matrix (the static transition corresponding to RSB while the dynamical transition
being given by the so-called ``marginality condition'').
\section{The limiting distribution of the rescaled energy levels.} 
\label{limiting_distribution}
\noindent
The knowledge of the eigenvalues of $J$ imposes simple bounds on the  
energy of any spin configuration. Indeed a state vector $\s $ can be 
decomposed into his projections on the various orthogonal eigenspaces 
relatives to different eigenvalues.  Here, due to orthogonality, the 
possible eigenvalues are $+1,-1$ so that 
\be
 -\frac{N}{2} \leq H (\s) \leq  \frac{N}{2}.
\ee
Let us consider the rescaled and shifted Hamiltonian 
(representing the energy per site, or energy density of the model, plus 
the `zero point'
energy $1/2$)
\be
h(\s) = {H(\s)\over N}+{1\over 2}
\ee
which takes values in $[0,1]$. We shall show that in the limit $N\to \infty$ the energy density $h$
gets $\delta$-distributed at $x=1/2$. We point out that this property can be
immediately proved for the Curie-Weiss model, thus indicating a mean field
behaviour of the present model in the thermodynamic limit. To this end consider
the partition function $Z_N$ at inverse temperature $\beta$:
\be
\label{partitionfct}
Z_N(\beta) = \sum_{\s \in \S_N} \exp{(-\beta H(\s))}=2^N\,\E_N (e^{-\beta H}),
\ee
where $\E_N$ denotes the expectation wrt $\P_N$, and note that the characteristic
function of
$h$ can be written as
\be
\label{nrelazione}
\E_N(e^{-\lambda h}) = e^{-\lambda/2}\, {Z_N(\lambda/N)\over 2^N}.
\ee
This expression will prove useful to compute the limiting expression of the
characteristic function of the energy density $h$ without knowing the 
expression of all its moments. 
To see this, we first decouple the spins as follows: let $B$ be an
orthogonal matrix such that
$B^TJB=D$ with
$D={\rm diag}\, (d_1,\dots ,d_N)$. Since ${\rm det}\, J\ne 0$ we have 
$d_i\neq 0$, $i=1,\dots
,N$, and 
${\rm det}\, J^{-1}=\prod_{i}d_i^{-1}$. 
Let $u\in \R^N$ be such that $\s = B u$.
We have
$<J\s,\s>=<Bu,JBu>=<u,Du>$, and thus
\begin{eqnarray}
\exp({\lambda\over 2N}<J\s,\s>)&=&\prod_{i=1}^N\exp({\lambda\over 
2N}d_iu_i^2 )
\nonumber \\
&=&\prod_{i=1}^N{1\over
\sqrt{2\pi}}
\int_{-\infty}^\infty \exp\left(-{x_i^2\over 2}+
\sqrt{{\lambda d_i\over N}}\,\, u_ix_i\right) dx_i \nonumber \\
&=&{1\over (2\pi)^{N/2}}\int_{\R^N}
\exp\left(-{1\over 2}<x,x>+
\sqrt{{\lambda \over N}}\,\, <u, D^{1/2}x>\right) \, dx\nonumber \\
&=&{{\rm det}\, J^{-{1\over 2}}\over (2\pi \lambda)^{N/2}}\int_{\R^N}
\exp\left(-{1\over 2\lambda}<y,J^{-1}y>+
\, <\s, {y\over \sqrt{N}}>\right) \, dy,\nonumber 
\end{eqnarray}
which, together with (\ref{hamiltonian}), (\ref{partitionfct}) and 
(\ref{nrelazione}) yields
\be
\E_N(e^{-\lambda h}) = e^{-{\lambda/2}}\,
{{\rm det}\, J^{-{1\over 2}}\over (2\pi \lambda)^{N/2}}\int_{\R^N}
\exp \biggl(-{1\over 2\lambda}<y,J^{-1}y>+ 
\sum_i \log \cosh {y_i\over \sqrt{N}}\biggr) \, dy \nonumber
\ee
(As usual, the square roots appearing in the above formulas are only
apparently ill defined: they disappear in the expansion because it contains
only the even terms). The above integral can be evaluated by means of
 standard
high-temperature expansion techniques which turn out to be 
 considerably
simpler if one assumes that $\sum_{i}J_{ii}=0$ (see \cite{PP}). As we have
already 
 noted,
 this assumption amounts to fix at zero the mean value of the
energy. 
 Also, the division by $N$ of the argument of the
partition function leads to a convergence domain which is increasing as $N$ 
itself.
In this way, the asymptotic expression (for $N\to \infty$) of 
$\E_N(e^{-\lambda h})$
can be written in the form 
\be
\E_N(e^{-\lambda h})=e^{-\lambda/ 2}\, e^{NG(\lambda/N)}\; 
\left(1+{\cal O}
(N^{-1}) \right)
\ee
where the function $G(x)$ is an effective specific free energy. 
For the orthogonal
interaction matrix (\ref{sine}) one finds \cite{PP}:
\be
G(x) = {1\over 4}\left[ \sqrt{1+4x^2}-\log\left({1+\sqrt{1+4x^2}\over 
2}\right)-1\right].
\ee
It has the following expansion in the
vicinity of $x=0$:
\be
G(x) = {x^2\over 4} + {\cal O}(x^3),
\ee
which, by the way, coincides with what one obtains for the SK model.
This yields
\be
e^{NG(\lambda/N)} = 1+{\lambda^2\over 4 N}+{\cal O}
\left(\lambda^3\over N^2\right).
\ee
Summarizing, we have found that for
any fixed $\lambda$, 
\be
\E_N(e^{-\lambda h})= e^{-\lambda/2}\,\left[1+{\cal O}
\left(1\over N\right)\right],\qquad N\to
\infty.
\ee
Using a well known theorem of probability theory \cite{Si} which 
says that a distribution function $G_N$ converges weakly to $G$ if
and only if 
$\varphi_N(\lambda) \to \varphi (\lambda)$ for any $\lambda$ (where 
$\varphi_N(\lambda)$ and
$\varphi (\lambda)$ are the characteristic fcts of $G_N$ and $G$ 
respectively)
and noting
that 
$\varphi (\lambda) = e^{-\lambda/2}$ is the characteristic function of 
the distribution fct
$G(x)={\chi}_{[{1\over 2},\infty)}$, we then conclude that the 
distribution of $h$ tends
to ${\chi}_{[{1\over 2},\infty)}$.
\vsni
\noindent
\noindent
\section{Flipping spins from the ground state and statistics of levels}
\label{flipping}
\begin{figure}[ht]
\includegraphics[width=2.7in]{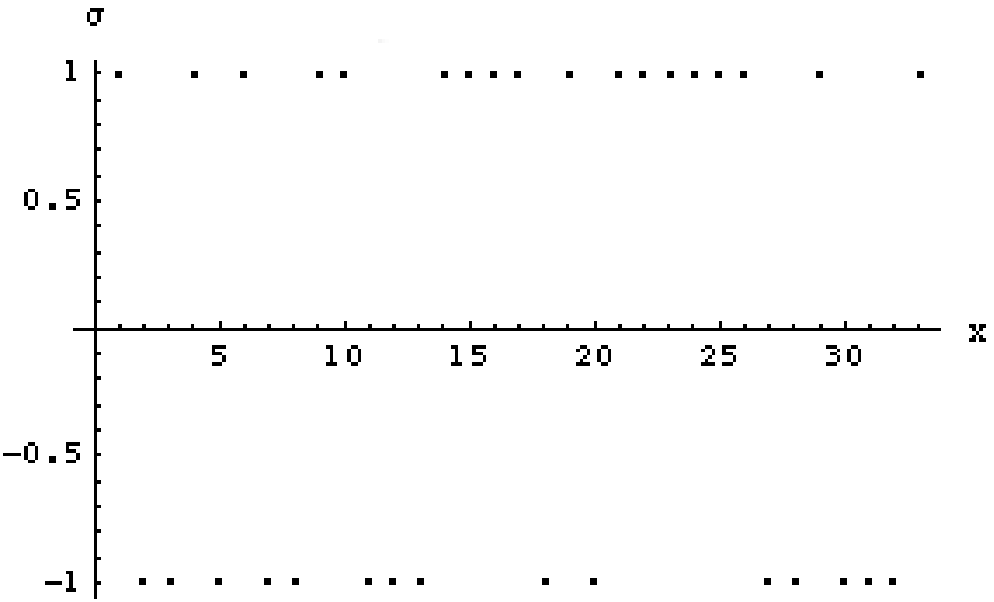}
\includegraphics[width=2.7in]{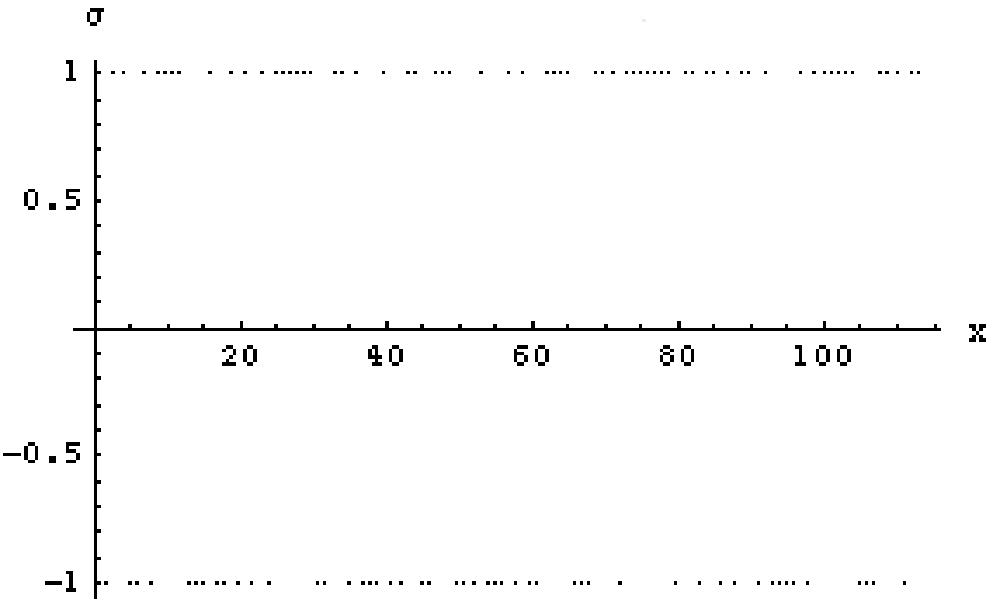}
\caption{\baselineskip=13pt {\small {Ground state for $N=33$ and 
$N=113$}}}
\protect\label{ground1}
\end{figure}

\begin{figure}[ht]
\includegraphics[width=5in]{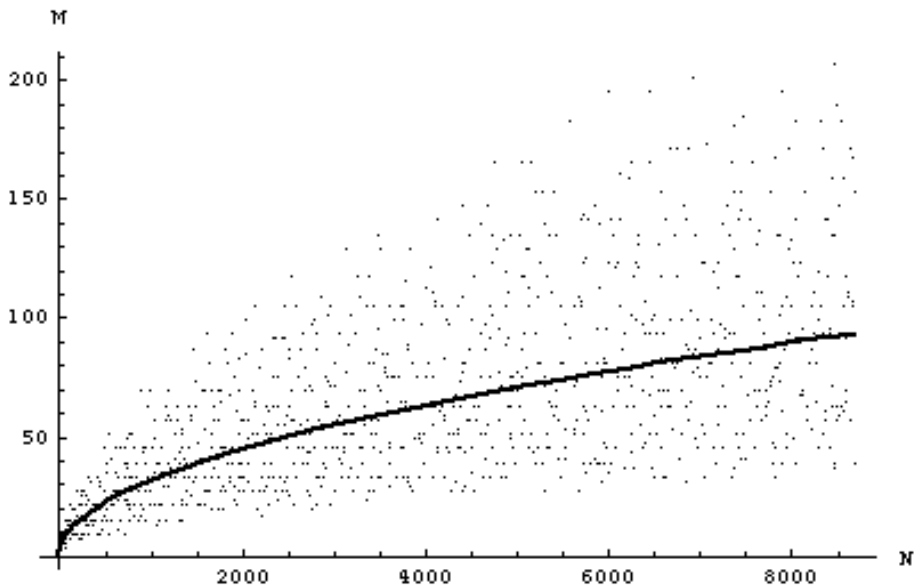}
\caption{\baselineskip=13pt {\small {Total magnetization $N m^0$ of the 
ground state versus $N$, $2N+1=4m+3$ prime. The continuous curve is 
$\sqrt{N}$.}}}
\protect\label{magn1}
\end{figure}

As already noted in  \cite{MPR}, for special values of $N$ the  ground state, 
i.e. the  configuration 
$\s^0 \in \S_N$ which minimizes the energy, can be 
explicitly constructed.
Indeed,
for $N$ odd such that $p = 2N+1$ is prime of the form $4m+3$, where 
$m$ is
an integer, let $\s^0$ be the state given
by the  sequence of Legendre symbols, i.e.
\be
\s^{0}_x = \left (\frac{x}{p}\right ) = \left\{ \begin{array}{ll}
                                            +1, & \mbox{if $x=k^2 
(\bmod p)$}, \\
                                            -1, & \mbox{if $x \neq k^2 
(\bmod p)$},        
                                             \end{array}
                                     \right.
\ee
with $k=1,2,\ldots,p-1$. Then (see the Appendix):
\be \label{gs}
H(\s^0)=-{N\over 2}\cdot
\ee
A typical ground state for $p$ prime of the form $4m+3$ reflects the 
well 
known random distribution of the Legendre symbols (see Fig. 
\ref{ground1}, 
where a pair of ground states are shown for two different $N$ values).
No structure is present at any scale. Nevertheless, denoting by 
$m^0$ the  {\em specific magnetization} of the ground state, i.e.
\be
m^0\,=\,\frac{1}{N}\sum_{x=1}^{N}\s^0_x\,
     =\,\frac{1}{N}\sum_{x=1}^{N}\legendre{x}{p}\, ,
\ee
one observes that it tends to be a positive function of $N$, fluctuating
around the value
$1/\sqrt{N}$. To let the reader better appreciate this fact we plot in Fig.
\ref{magn1}  the total 
magnetization $Nm^0$ versus $N$.

\noindent
For the remaining part of the paper we will restrict to $p=2N+1$ 
prime, with $p=3\,(\mbox{mod} 4)$. We point out that this set has measure zero as a
subset of the natural numbers. However, we have strong numerical evidence that 
some relevant properties that we are going to discuss hereafter, such as the
behaviour of
$m^0(N)$, the statistics of energy levels and the number of metastable
states (see below) are somehow generic in $N$. 

\begin{figure}[ht]
\includegraphics[width=4.5in, angle=-90]{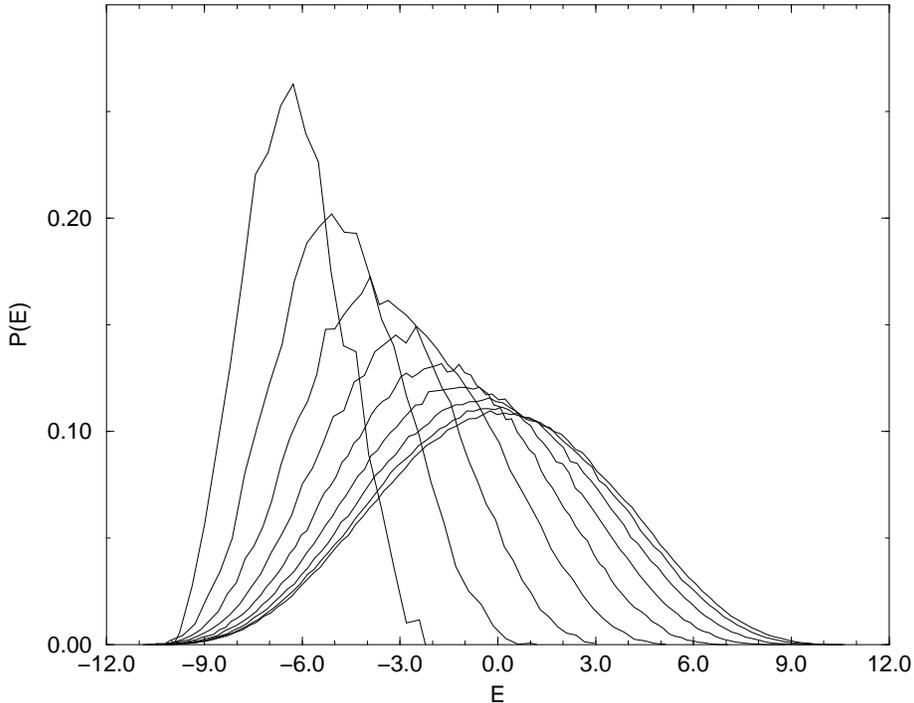}
\caption{\baselineskip=13pt {\small {Distributions of the energy over the
subspaces $\Omega_s$. The size of the system is $N=23$ and $s=3,4,\ldots, 11$
from left to right}}}
\protect\label{overlap}
\end{figure}

\vsni
\noindent
Let $\Omega_{s} \subset \S_N$ be the subspace
consisting of the ${N \choose s}$ configurations obtained by starting 
from the
ground state $\s^0$ described above and flipping exactly $s$ different
spins. 
Each point of $\Omega_s$ can thus be identified with a $s$-dimensional 
vector
$\tau \in \{1,\dots ,N\}^s$ of the form $\tau=(x_1,\dots ,x_s)$, with 
$x_i\ne x_j$ for $i\ne j$, which
specifies the positions of the flipped spins along the chain of length 
$N$.
We then define the `flipping' map $L_{\tau}:\S_N \to \S_N$ as:
\be
\left(L_{\tau}\s\right)_x = \left\{ \begin{array}{ll}-\s_x, & x \in 
{\tau}, \\
                          \;\;\,\s_x, & x \notin {\tau}. 
\end{array}\right.
\ee 
In this way we can write 
\be
\Omega_s = \{L_{\tau}\s^0\}_\tau.
\ee
The correspondence $\tau \to \sigma$ given by $\s = L_\tau \s^0$ is 
plainly one-to-one.  Therefore in the sequel we shall freely identify a state
$\s = L_\tau \s^0$ with the vector $\tau$.
Alternatively, we can proceed as follows. Define the {\sl
overlap} $q(\s)$ of a given configuration $\s\in \S_N$ with respect to 
the ground state $\s^0$ as:
\be
q(\s) = {1\over N} \sum_{x=1}^N \s_x
\s_x^{0},
\ee
so that $q(\s^0)=1$. Then 
\be
\Omega_s = \{ \s\in \S_N \, : \, q(\s) = 1-2{s\over N} \}
\ee
The following straightforward calculation yields the energy values on the space
$\Omega_s$: using the definition of $L_{\tau}$, the symmetry  of $J$
and the fact that the ground state $\s^0$ is an eigenvector of $J$ to 
the eigenvalue $1$
we have: 
\begin{eqnarray}\label{energyOmegas}
H(L_{\tau}\s^0 ) & = & -\frac{1}{2}\sum_{x\in \tau}\sum_{y\in \tau}
                   J_{xy}\s^0_x\s^0_y 
                   +\frac{1}{2}\sum_{x\in \tau}\sum_{y\notin \tau}
                   J_{xy}\s^0_x\s^0_y\nonumber \\
             &   & +\frac{1}{2}\sum_{x\notin \tau}\sum_{y\in \tau}
                   J_{xy}\s^0_x\s^0_y
                   -\frac{1}{2}\sum_{x\notin \tau}\sum_{y\notin \tau} 
                   J_{xy}\s^0_x\s^0_y \nonumber \\
             & = & -\frac{1}{2}\sum_{x,y=1}^{N}J_{xy}\s^0_x\s^0_y 
                   +2\sum_{x\in \tau}\sum_{y\notin \tau}
                   J_{xy}\s^0_x\s^0_y \nonumber \\
             & = & -\frac{N}{2} 
                   +2\sum_{x\in\tau}\sum_{y=1}^{N}J_{xy}\s^0_x\s^0_y 
                   -2\sum_{x\in\tau}\sum_{y\in\tau}J_{xy}\s^0_x\s^0_y 
\nonumber \\
             & = & -\frac{N}{2} 
                   +2\sum_{x\in\tau}(\s^0_x)^2 
                   -2\sum_{x\in\tau}\sum_{y\in\tau}J_{xy}\s^0_x\s^0_y 
\nonumber \\
             & = & -\frac{N}{2} + 2s - 2\sum_{x\in \tau}\sum_{y\in 
\tau}
                   J_{xy}\s^0_x\s^0_y.       
\end{eqnarray}
\noindent
Notice that for the Ising mean-field interaction: $J_{xy}=1/N$, one 
finds 
\be
H(L_{\tau}\s^0) = -\frac{N}{2} + 2s - 2\frac{s^2}{N} = -{N\over 
2}\left(1-{2s\over N}\right)^2 .
\ee
It is now possible to study the distribution of the energy levels on 
the individual subspaces $\Omega_s$, where $s=0,1,\ldots,N$.  
Let $p_{s}$ be the probability distribution restricted on 
$\Omega_s$, i.e.
\be
\label{ps}
p_{s}(C)={{N \choose s}}^{-1} \sum_{\s \in C \cap \Omega_s} 1,
\ee
and let $\E_s$ denote the expectation wrt $p_s$.
The $n$-th moment of the energy $H$
on the subspace $\Omega_s$ is given by
\be
\E_s (H^n) \equiv \int_{\Omega_s}H^{n} (\s)\, dp_{s}(\s) 
={{N \choose s}}^{-1} \sum_{\tau\in\Omega_s}H^n(L_{\tau}\s^0 ),
\ee
so that the $n$-th moment $\E_N(H^n)$ of the energy on the whole configuration
space
$\Sigma_N$ is
\be
\E_N(H^n)\equiv\int_{\S_N}H^n(\s)\, d\P_N(\s)={1\over 2^N}\sum_{s=0}^{N}{N
\choose s}\,\E_s (H^n).
\ee
A tedious but straightforward calculation (see  Appendix) yields the
following expressions for the first two $s$-moments: 
\be
\label{moment1}
\E_s (H)=\, -{N\over 2}\left(1-{2s\over N}\right)^2,\qquad 
\ee
\be
\label{moment2}
\sigma_s^2(H) \equiv
\E_s (H^2) - (\E_s (H))^2  = 
\frac{4s(s-N)(2s^2-2sN+N)}{N^2(N-2)}
\ee
and consequently
\be
\E_N (H) = -{1\over 2},\qquad
\E_N (H^2) - (\E_N (H))^2= {N-1\over 2}.
\ee
These results indicate that, at variance with the 
ferromagnetic case where the energy is constant on each subspace $\Omega_s$, here
there is a significant overlap between the distributions (for different $s$
values) of the energy when restricted to $\Omega_s$.
In particular, from the espression of $\sigma_s^2$ we see that there can be a
large number of states having small overlap with the ground state but
nevertheless with energy very close to $-N/2$. For example we have
$\sigma^2_{N/2}\simeq N/2$, indicating that the energy restricted to the subspace
$\Omega_{N/2}$ may fluctuate over the whole energy range. 
This simple phenomenon is intimately related to the existence of metastable 
states and it will prove crucial in the understanding of  the zero temperature
dynamics, as discussed below. Fig. \ref{overlap} shows the distributions of the 
energy restricted to various subspaces $\Omega_s$.

\vskip 0.2cm
\noindent
Another quantity of interest is the specific magnetization 
$m(\s)$ of an arbitrary state $\s \in \S_N$, given by:
\be
m(\s) = {1\over N} \sum_{x=1}^N \s_x,
\ee
In particular, given $\tau=(x_1,\dots ,x_s)$, we have
\be
m(L_\tau\sigma^0)\,\equiv m(\tau)=\, m^0
-\frac{2}{N}\sum_{x\in\tau}\legendre{x}{p}.
\ee
Clearly,
\be
\frac{1}{\binomial{N}{s}}\sum_{\tau\in\Omega_s}\left(\sum_{x\in\tau}\legendre{x}{p}\right)\,=\, \frac{1}{\binomial{N}{s}}\sum_{x=1}^{N}\binomial{n-1}{s-1}\cdot\legendre{x}{p}=\, s\cdot m^0\, ,
\ee
i.e.
\be
E_s(m)\,=\, \frac{1}{\binomial{N}{s}}\sum_{\tau\in\Omega_s} m(\tau)\,=\, m^0
\left(1-2{s\over N}\right).
\ee
Moreover, we show in the Appendix that
\be\label{variancemagn}
\sigma_s^2(m)\equiv
E_s(m^2)-(E_s(m))^2=\frac{4s(N-s)}{N^3(N-1)}+
 \frac{4 s (s-1)}{N^3(N-1)}\cdot
\sum_{x=1}^{p}\legendre{x}{p} d_N(x),
\ee
where $d_N(x)$ is the integer valued function giving the number of elements $u\in
\{1,\dots, N\}$ such that $u^{-1}x\in \{1,\dots, N\}$ (where $u^{-1}$ denotes the
inverse $\mod p$ of $u$). As will be discussed in the Appendix, $d_N(x)$ takes
values around $p/4$ with rather small fluctuations. Since
$\sum_{x=1}^{p}\legendre{x}{p}=0$, the last term in (\ref{variancemagn}) can
beconsidered as a small correction to the constant value $4s(N-s)/N^3(N-1)$.

\section{Zero temperature dynamics and metastable states.}
We first introduce the following discrete $1-$flip dynamics, given by:
$$
\s (t+1)=\left\{\begin{array}{l}L_{\omega(t)}\, 
\,\sigma(t),\qquad\mbox{if
$H(L_\omega\s)<H(\s)$,}\\\ \qquad 
\s(t),\qquad\mbox{otherwhise,}\end{array}\right.
$$
where, for each $t$, $\omega (t)$ is chosen randomly in 
$\{1,\ldots,N\}$ with uniform
distribution. Choosing an initial
condition $\s(0)$ at random with respect to $\P_N$, one obtains a 
random orbit
$\{\s(0),\s(1),\ldots,\s(\ell)\}$ for any realization $\{\omega 
(t)\}_{1\leq
t\leq \ell}$ of length
$\ell$.
As a consequence of the previous analysis, we have the following 
remarks.
\begin{itemize}
\item On one hand, it may happen that starting from $\s(0)$ one reaches 
after $t$ iterations a state $\s(t)\in
\Omega_s$, of the form
$\s(t)=L_\tau \s^0$ for some $\tau =(x_1,\dots ,x_s)$,
such that $H(L_\tau \s^0) < H(L_\omega L_{\tau} \s^0)$ for any 
$\omega\in \{x_1,\dots ,x_s\}$.
\item On the other hand one can reach $\s(t)\in\Omega_s$ such that for 
some $\omega \in \{1,\dots ,N\}$,
$L_{\omega}\s\in\Omega_{s+1}$ and $H(L_\omega\s (t))<H(\s(t))$. 
\end{itemize}
Due to the above observations, the overlap function $q(\s(t))$ is in 
general not monotonically non-decreasing
along a given random orbit (this at variance with the Ising mean field 
model). In particular there might
be metastable states \cite{NS}.
Now, given $\omega\in\{1,\dots ,N\}$, we shall say that a configuration 
$\s\in\Sigma_N$ is {\it $\omega$-stable}
if
\be
H(L_\omega\s)\,> \,H(\s).
\ee
Moreover, we say that $\s$ is
{\it $1$-flip stable}  (or {\it metastable}) if it is $\omega$-stable 
$\forall\omega\in\{1,\ldots,N\}$.\par
\begin{figure}[ht]
\includegraphics[width=5.5in, angle=-90]{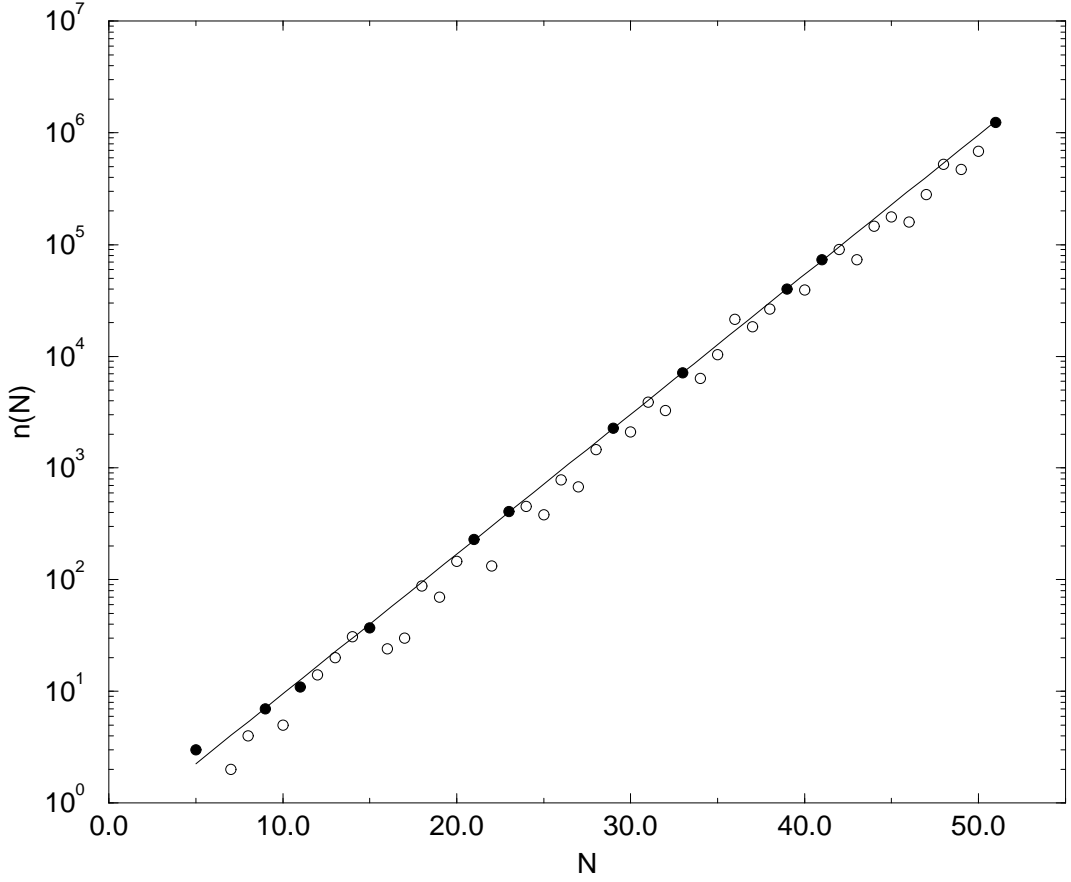}
\caption{\baselineskip=13pt {\small { 
The number of metastable states $n(N)$ for $N$.
The line represents the best fit $n(N)\approx e^{\lambda N}$ 
with $\lambda\approx 0.28$ for values of $N$ such
that $p=2N+1$ is a prime of the form $4m+3$ (filled points).
Other points are for generic integer $N$. }}}
\protect\label{metastate}
\end{figure}

\noindent
We denote by $n(N)$ the total number of such metastable states as a 
function of  $N$. From (\ref{energyOmegas}) one readily obtains that
\be
\label{difference}
H(L_\omega \s)\,=\,H(\s)\,+
\,2\,\sum_{x\neq\omega}\,J_{x\omega}\,\s_x\, \s_\omega,
\ee
so that $\s$ is $\omega$-stable if and only if
\be \label{omega}
\left(J\s\right)_\omega \s_\omega > J_{\omega \omega}.
\ee
Summing over $\omega$ and using (\ref{orth}) we see that if $\s$ is 
$1$-flip stable
then
\be \label{1-flip}
<J\s , \s> \; > \; 1.
\ee
Recalling the expression 
(\ref{hamiltonian}) of the Hamiltonian we see that a necessary condition for
$\s\in \S_N$ to be $1$-flip stable is that
\be
H(\s) < \E_N(H)=-{1\over 2}\, \cdot
\ee
The main goal of this paper is to give an estimate of the number of 
metastable states for any given $N$. To this end we first performed 
some numerical investigations. For $N\leq 30$ we performed an exact 
enumeration of all configurations, whereas for larger $N$ we
run the zero temperature dynamics described above (``deep quench'')
for a number or realizations $\{\omega  (t)\}$ as large as $10^8$ for
bigger sizes, keeping track of the metastable states.
As shown in Fig.
\ref{metastate}, the growth of these states is exponential for
generic values of the $N$. 
The best numerical fit yields
\be
n(N)\simeq C\cdot e^{0.28 N}. 
\label{ntrue}
\ee
We remark that the same behaviour has been observed
in \cite{PP} for the Random Orthogonal Model; for spin glasses see \cite{PP2}. 
We now proceed to give a partial justification of this result by means of probabilistic
arguments. 

\begin{figure}[ht]
\includegraphics[width=2.7in, angle=-90]{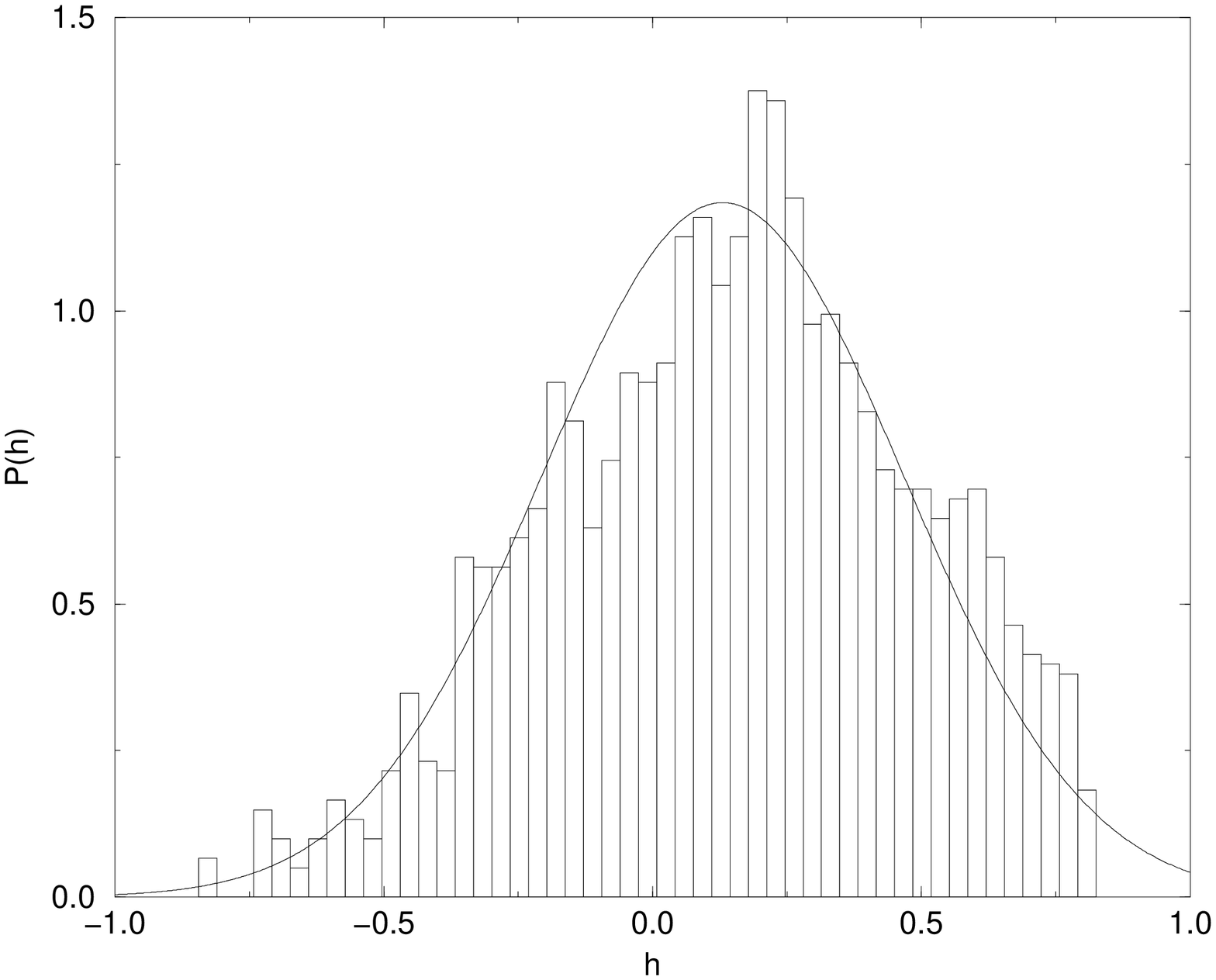}
\includegraphics[width=2.7in, angle=-90]{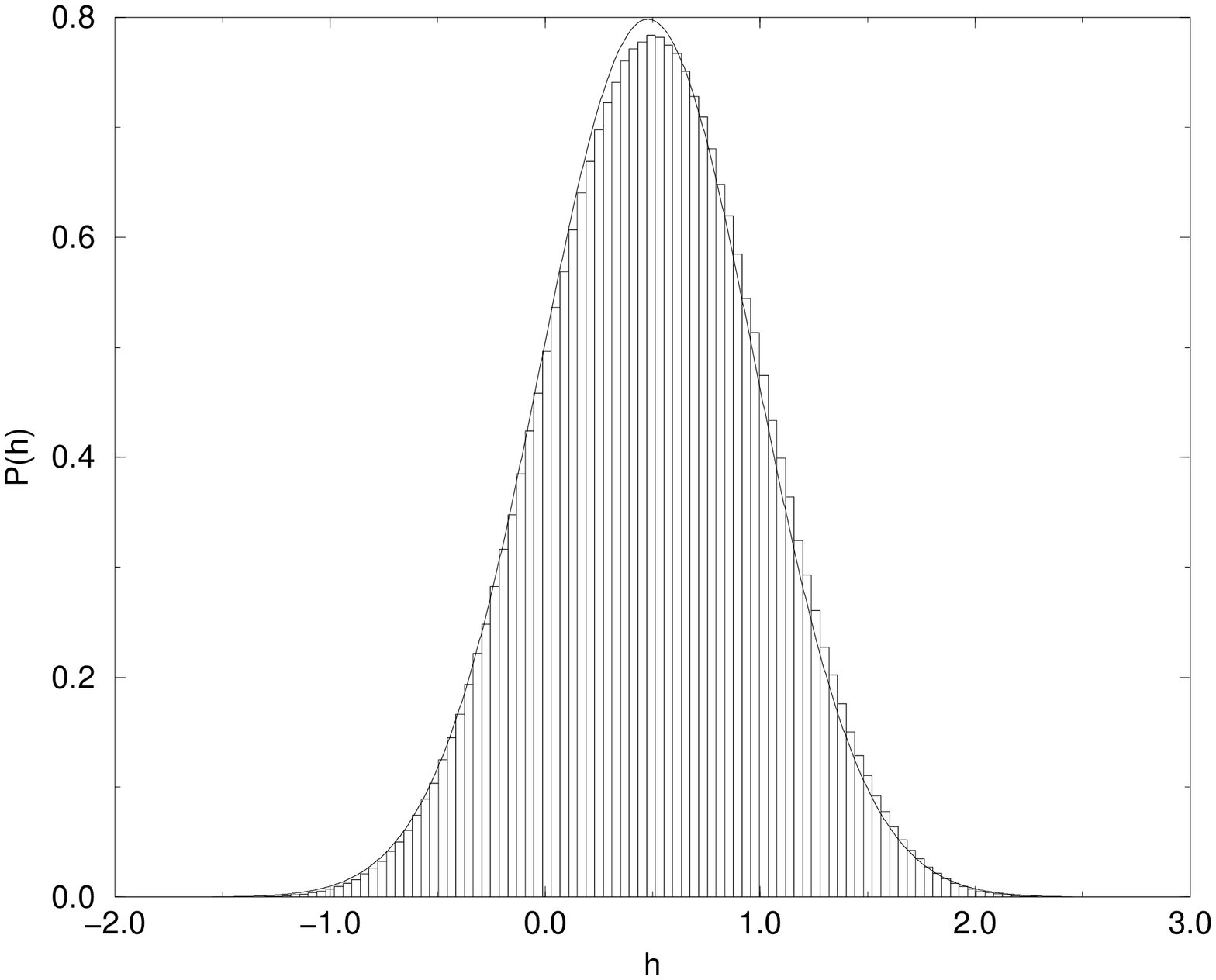}
\caption{\baselineskip=13pt {\small {
Distribution of the function $h(\tau,\omega)$ for a fixed 
$\omega\in\{1,\dots ,N\}$ and $\tau$ varying in $\Omega_s$. 
Here $N = 23$ and $s=3$ (left) and $s=11$ (right). The solid 
line is the gaussian distribution in the r.h.s. of 
(\ref{gauss}). }}}
\protect\label{h}
\end{figure}

\noindent
Let $\tau=(x_1,\dots ,x_s)$ and $\omega \in \{1,\dots ,N\}$ be 
given. 
Using (\ref{difference}) and $J\sigma^0\,=\,\sigma^0$, it is easy to 
see that if
$\omega\notin\tau$  (i.e. $L_\omega\s \in\Omega_{s+1}$)
$$
H(L_\omega\s)\,=\,H(\s)\,+\,2\,
\left(1- J_{\omega\omega} - 2\, 
\sum_{x\in\tau}\,J_{x\omega}\,\sigma^0_x\,\sigma^0_\omega 
\right),
$$
whereas, if $\omega\in\tau$ (i.e. $L_\omega \s \in\Omega_{s-1}$), we 
have
$$
H(L_\omega\s)\,=\,H(\s)\,-\,2\,\left(1+ J_{\omega\omega}
-2\,\sum_{x\in\tau}\,J_{x\omega}\,\sigma^0_x\,\sigma^0_\omega \right). 
$$
If we define
\be
h(\tau,\omega)\,= \sum_{x\in \tau}
J_{x\omega}\,\sigma^0_x\s^0_\omega,
\ee
we then see that  a configuration $\s=L_\tau\sigma^0$ 
is $\omega$-stable
if and only if 
\be \label{omegastable}
\left\{\begin{array}{lll}
h(\tau,\omega)&<& {1\over 2}(1-J_{\omega \omega}),\qquad \mbox{\rm if 
}\omega\notin\tau,\\
\\
h(\tau,\omega)&>&{1\over 2}(1+J_{\omega \omega}),\qquad \mbox{\rm if 
}\omega\in\tau.
\end{array}\right.
\ee
We now dwell upon the problem of characterising the behaviour of the
function $h(\tau,\omega)$ so as the condition (\ref{omegastable}) can be effectively used
to estimate the number of metasable states. 
Let us rewrite $h(\tau, \omega)$ in the form
\be
h(\tau,\omega) =\frac{2}{\sqrt{p}}\sum_{x\in\tau} \, \xi_x (\omega),
\ee
where
\be
\label{xi}
\xi_x(\omega) :=\left (\frac{\omega x}{p}\right )\,\,\sin
\left(\frac{2\pi \omega x}{p}\right).
\ee
Now, having fixed $\tau$ and $x\in \tau$, we can view the function 
$\xi_x (\omega)$ defined in (\ref{xi}) as a random
variable uniformly distributed on $\{1,\dots, N\}$ and taking values in 
$[-1,1]$. Its
mean $\mu_x$ and variance $\sigma^2_x$ are easily computed:
\be
\mu_x={1\over N}\sum_{\omega =1}^N \left (\frac{\omega x}{p}\right 
)\,\,\sin
\left(\frac{2\pi \omega x}{p}\right) ={\sqrt{p}\over 2N},
\ee
and, using (\ref{orth}),
\be
\sigma^2_x = {1\over N}\sum_{\omega =1}^N  \,\sin^2
\left(\frac{2\pi \omega x}{p}\right) - \mu^2_x = {p\over 
4N}\left(1-{1\over N}\right).
\ee
Here we want to study the behaviour of the sum
$\eta (\tau, \omega) :=\sum_{x \in\tau}\xi_x (\omega)$. We remark that 
this sum, and thus 
$h(\tau, \omega)=2\eta (\tau, \omega)/\sqrt{p}$, has
to be regarded as a r.v. defined on the product of two probability 
spaces: for each fixed $\tau$, it is
the sum of the i.i.d.r.v.'s $\xi_x (\omega)$ on the space $\{1,\dots 
,N\}$ with uniform distribution
(this comes from the very definition of the zero temperature dynamics); 
on the other hand, for each fixed 
$\omega$, it can be regarded as a r.v. on $\Omega_s$ viewed as a
probability space endowed with the distribution $p_s$. Its mean is
given by (recall that the symbol $\E_s$ denotes the expectation wrt $p_s$):
\begin{eqnarray}
\E_s \eta &=& {{N \choose s}}^{-1} \sum_{\tau = (x_1,\dots, x_s)} 
\sum_{x\in \tau}\xi_x (\omega)
= {{N \choose s}}^{-1} \sum_{x =1}^N {N-1 \choose s-1} \xi_x (\omega)\\
&=& {s\over N} \sum_{x =1}^N \xi_x (\omega)={s\sqrt{p}\over 
2N}.\nonumber
\end{eqnarray}
which does not depend on $\omega$ and equals $s$ times $\mu_x$. Along 
the same lines one shows that
\be
{\rm Var}_s \eta = {sp\over 4 N}\left(1-{s\over N}\right).
\ee
Notice that unlike the means, here we have $ {\rm Var}_s \eta \ne s 
\cdot \sigma_x^2$.
This discrepancy comes from the fact that, for any fixed $\omega$,
the sequence $\xi_{x_1},\xi_{x_2},\dots ,\xi_{x_s}$ is a sequence of distinct (and
ordered) elements so that by no means we can view $\eta$ as a sum of independent and
identically distributed objects.
Nonetheless, also supported by strong numerical evidence (see Fig. \ref{h}), we claim
that a version of the central limit theorem is applicable so that when $N\to \infty$,
$s\to
\infty$ with $s/N\to \lambda$, we have
\be
\label{normal}
p_s\left( \alpha < {\eta - \E_s \eta\over\, \sqrt{{\rm Var}_s \eta} } < 
\beta \right)
\rightarrow
{1\over \sqrt{2\pi}} \int_{\alpha}^\beta e^{-y^2/2}\, dy.
\ee
Assuming the validity of (\ref{normal}), performing the change of variables 
$y=(x-\lambda)/\sqrt{\gamma}$, with $\gamma = \lambda(1-\lambda)$, and setting
$a=\lambda + \alpha \sqrt{\gamma}$, 
$b=\lambda + \beta \sqrt{\gamma}$, we thus obtain an asymptotic gaussian
distribution for $h(\, \cdot \, ,\omega)$:
\be\label{gauss}
p_s\left( a< h(\tau,\omega)  < b \right) \rightarrow
{1\over \sqrt{2\pi\gamma}} \int_{a}^b e^{-(x-\lambda)^2/2\gamma}\, dx.
\ee
Note that the r.h.s. does not depend on $\omega$. One can actually say more: for any
${\tilde\omega},\omega\in\{1,\ldots,N\}$ we have
$h({\tilde\omega}^{-1}\tau,{\tilde\omega}\omega)=h(\tau,\omega)$.
Therefore the set of values of $h(\,\cdot \, ,\omega)$ on $\Omega_s$ does not depend
on the choice of $\omega$, i.e.
$\left\{h(\tau,\omega)\right\}_{\tau\in\Omega_s}\,=
\,\left\{h(\tau',\omega')\right\}_{\tau'\in\Omega_s}$, for all $\omega$,
$\omega'$.
\vskip 0.2cm
\noindent
Having fixed an order for the lattice points
$(\omega_1,\ldots,\omega_N)$, $\omega_j\neq\omega_k$,
we now consider the following quantities:
\be
\pi_{s,N}(\omega_k) = p_s \left( \left\{ \sigma \in \O_s \, : \, 
H(L_{\omega_k}\s)\,> \,H(\s)
\right\}\right),
\label{pomega}
\ee
the $p_s$-probability that a randomly chosen state $\s \in \O_s$ is
$\omega_k$-stable,
\begin{eqnarray}
\pi_{s,N}(\omega_{k+1}\vert \,\omega_1 ,\ldots ,\omega_k)  
 & = & p_s \left( \left\{  \sigma \in \O_s \, : \, 
       H(L_{\omega_{k+1}}\s)\,> \,H(\s) \right. \right. \nonumber\\
 &   & \;\;\;\;\;\;\;\left. \left. \vert \, H(L_{\omega_{j}}\s)\,> 
\,H(\s), \; 
       \omega_j = \omega_1,\ldots\omega_k \right\}\right),
\end{eqnarray}
the conditional $p_s$-probability that a randomly chosen state 
$\s 
\in \O_s$ is $\omega_{k+1}$-stable given that it is 
$\omega_j$-stable
for $j=1,\ldots,k$, and 
\be
\pi_{s,N} = p_s \left( \left\{ \sigma \in \O_s \, : \, H(L_\omega\s)\,> 
\,H(\s),\;
\omega = \omega_1,\dots, \omega_N \right\}\right),
\ee
the $p_s$-probability that a randomly chosen state $\s \in \O_s$ is
$1$-flip stable (i.e. stable for all possible flipping). Notice that by 
condition (\ref{omegastable}) the last quantity can be written as

\be \label{right}
\pi_{s,N} ={N \choose s}^{-1}\sum_{(x_1,\dots, x_s)=\tau \atop 
(y_1,\dots, y_{N-s})=\tau^c}
\prod_{i=1}^s \theta \left(h(\tau,x_i)> {1+J_{x_i x_i}\over 2}\right) \, 
\prod_{j=1}^{N-s} \theta \left(h(\tau,y_j)< {1-J_{y_i y_i}\over 2}\right).
\ee

\noindent
The three quantities introduced above are related by the following 
identity:
\be
\label{prodotto}
\pi_{s,N}\,=\,\pi_s(\omega_1) \cdot 
              \pi_s(\omega_2\vert \omega_1) \cdot 
              \pi_s(\omega_3\vert\omega_1,\omega_2)\cdots
              \pi_s(\omega_N\vert\omega_{1},\ldots,\omega_{N-1})
\ee
and the total number of $1$-flip stable states in $\S_N$ is,
by definition,
\be
\label{number}
n(N) = \sum_{s=0}^{N}{N \choose s} \pi_{s,N}.
\ee
We shall study the quantity $n(N)$ in the thermodynamic limit:
$N\to\infty$, $s\to\infty$,
with $s/N \to \lambda $ and $0<\lambda < 1$.
In this regime we write $\pi_{s,N} \equiv \pi_\lambda$ and apply
Stirling's formula to obtain
\be \label{Stirling}
{N \choose s} \sim {e^{N F(\lambda)}\over \sqrt{2\pi N \lambda 
(1-\lambda)}} \, ,\quad\hbox{with}\quad
F(\lambda) = -\lambda \log {\lambda} - (1-\lambda)\log{(1-\lambda)}.
\ee
Note that $F(\lambda)$ is concave and symmetric around $\lambda=1/2$, 
with
$F(1/2)=\log 2$. In this way we get for $N$ large and $s \simeq \lambda N$ with
$\lambda$ ranging in the unit interval,
\be 
n(N) \simeq  \int_0^1 { \sqrt{N\over 2\pi \lambda (1-\lambda)}}\,
 \exp{\biggl[N\biggl(F(\lambda)+{\log \pi_\lambda \over N}
\biggr)\biggr]}\, d\lambda.
\label{approx}
\ee
\noindent
It thus remains to estimate the probability $\pi_\lambda$. 
Let us consider first the unconditioned probability (\ref{pomega}). 
According to (\ref{omegastable}) and the total probability formula
we have:
\begin{eqnarray}
\label{totalprob}
\pi_{s,N}(\omega_k)
 & = & {s\over N} \, p_s \left( h(\tau,\omega_k)>
       {1+J_{\omega_k \omega_k}\over 2}\;\bigg|\; 
       \tau \ni \omega_k \right) \nonumber \\
 & + & \left(1-{s\over N}\right) \, p_s\left( h(\tau,\omega_k)< 
       {1-J_{\omega_k\omega_k}\over 2}\;\bigg| \;
       \tau \not\ni \omega_k \right).
\end{eqnarray}
Here $s/N$ and $1-s/N$ are the probabilities that $\tau \ni \omega_k$ 
and 
$\tau \not\ni \omega_k$, respectively. 
We can easily compute the conditional expectations
\be  
\E (h | \tau \ni \omega_k ) = {N -1\choose s-1}^{-1}{2\over 
\sqrt{p}}\sum_{\tau \ni
\omega_k}\sum_{x\in \tau}\xi_x(\omega_k)={s-1\over N-1}
-\left({s-N\over N-1}\right) J_{\omega_k\omega_k}\, ,
\ee
and
\be
\E (h | \tau \not\ni \omega_k ) = {N -1\choose s}^{-1}{2\over 
\sqrt{p}}\sum_{\tau \not\ni
\omega_k}\sum_{x\in \tau}\xi_x(\omega_k)={s\over N-1}\left( 
1-J_{\omega_k\omega_k}\right)\,  .
\ee
In a similar way one can compute the variances $\gamma_+(\omega_k)$ and 
$\gamma_-(\omega_k)$ conditioned to the events
$\{\tau \ni \omega_k\}$ and $\{\tau \not\ni \omega_k\}$. 
For $N$ large and $s \simeq \lambda N$, retaining only terms ${\cal O}(1)$, one gets
\be
\E (h | \tau \ni \omega_k ) \simeq
\E (h | \tau \not \ni \omega_k ) \simeq 
\lambda,\qquad
\gamma_-(\omega_k) \simeq \gamma_+(\omega_k)\simeq \gamma =\lambda( 1- \lambda).
\ee
Moreover in the thermodynamic limit specified above we write $\pi_{s,N}(\omega_k)\equiv 
\pi_{\lambda}(\omega_k)$ 
and argue from (\ref{gauss}) the following approximate 
expression 
for $\pi_{\lambda}(\omega_k)$:
\begin{eqnarray}\label{erf}
\pi_{\lambda}(\omega_k)
 & \simeq & {\lambda\over\sqrt{2\pi\gamma}}
             \int_{1/2}^{\infty}
             e^{-(x-\lambda)^2/ 2\gamma}\, dx   + {(1-\lambda)\over\sqrt{2\pi\gamma}}  
             \int_{-\infty}^{1/2} 
             e^{-(x-\lambda)^2/ 2\gamma}\, dx \nonumber \\
& = & {1\over 2} +
\left ( {1\over 2} - \lambda \right )
{\rm erf}\left ( {{1\over 2} - \lambda \over\sqrt{2\gamma}} \right )
\end{eqnarray}
where we have denoted the error function by
\be
{\rm erf}(z) = {2\over\sqrt{\pi}}\int_0^z e^{-x^2}\, dx.
\ee  
It is not difficult to check that the r.h.s. of (\ref{erf}) is convex and symmetrix
around 
$\lambda = 1/2$, where it reaches its minimum value equal
to $1/2$.

\begin{figure}[ht]
\includegraphics[width=3.7in]{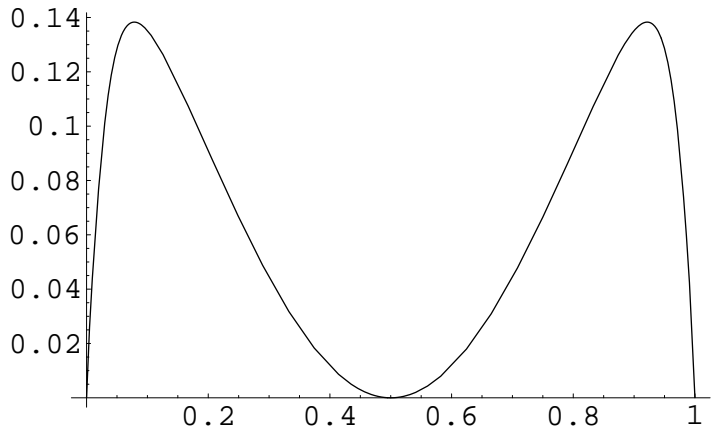}
\caption{\baselineskip=13pt {\small {
The function $G_1(\lambda)$ of equation (\ref{g1lambda}) for 
$\lambda\in [0,1]$. }}}
\protect\label{g1}
\end{figure}

\begin{figure}[ht]
\includegraphics[width=5.5in, angle=-90]{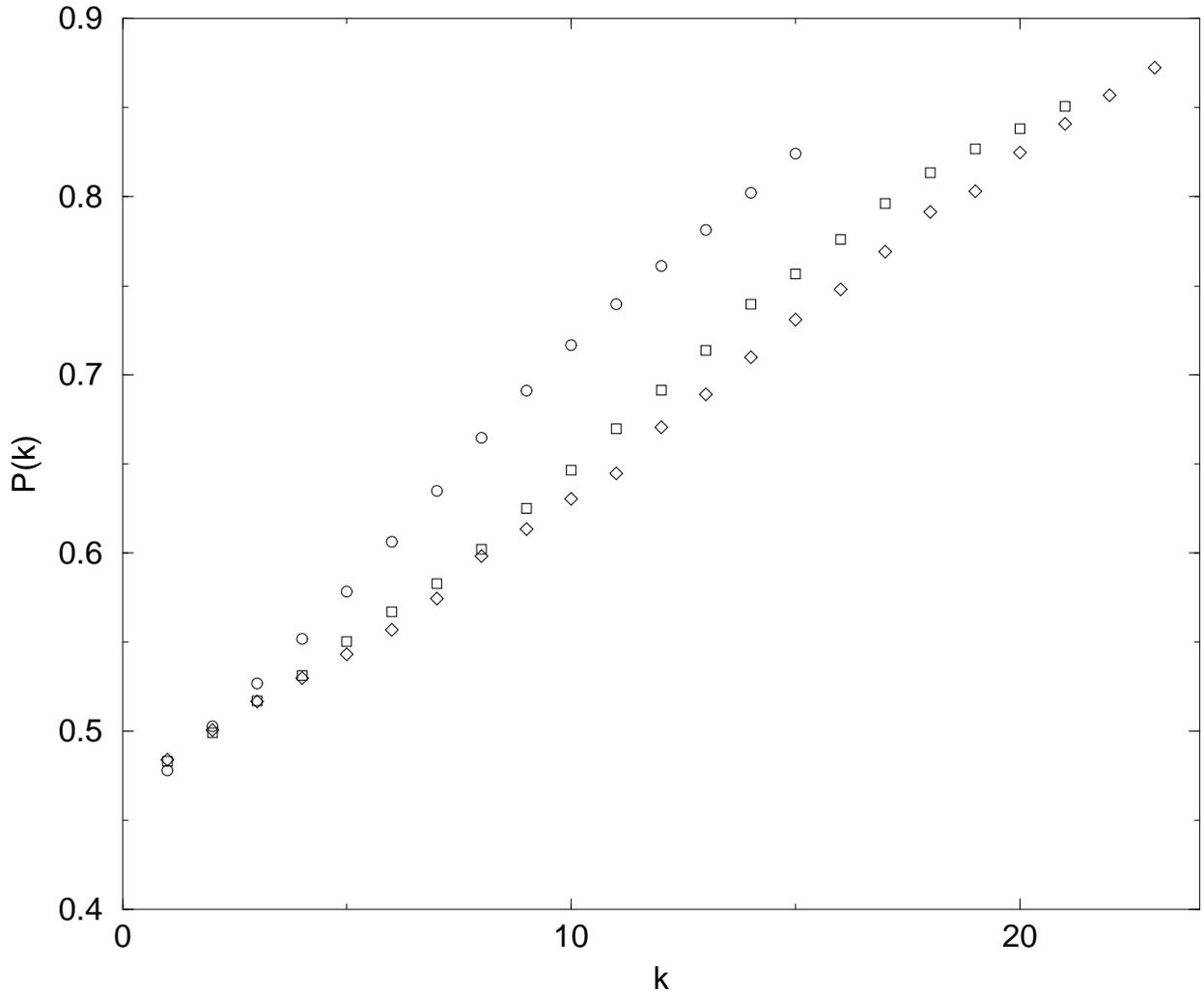}
\caption{  $P(k)$ versus $k$, for 
$N=15(\mbox{circle}),21(\mbox{square}),
23(\mbox{diamonds})$ and $s= [N/2]$. The conditional probabilities 
have 
been averaged over a sample of $10000$ choices of 
$(\omega_1,\ldots,\omega_N)$.}
\protect\label{conditional}
\end{figure}

\noindent
Let us now come to $\pi_{s,N}$. In principle this quantity is to be computed by specifying
the whole set of constraints embodied in (\ref{right}) or, which is the same, by computing
the conditional probabilities appearing in eq.(\ref{prodotto}). However, this appears to be
a difficult task.  A first approach which drastically simplifies this task is to forget
about the constraints implied by (\ref{right}) and assume that (in the thermodynamic limit)
the various
$\omega$-stability  conditions become mutually independent, that is
$\pi_{\lambda}(\omega_{k+1}\vert\omega_1,\ldots,\omega_k)=
\pi_{\lambda}(\omega_{k+1})$, for all $k=1,\dots ,N-1$, so that
\be
\pi_{\lambda}=\prod_{k=1}^N \pi_{\lambda}(\omega_k).
\ee
Recalling eq. (\ref{approx}) and (\ref{erf}), one is led to the following expression 
for $n(N)$:
\be \label{number1}
n(N) \simeq  \int_0^1 \sqrt{N\over 2\pi \lambda (1-\lambda)}\,
 \exp{\biggl(N G_1(\lambda)  \biggr)}\, 
d\lambda,
\ee
where 
\be
G_1(\lambda)=F(\lambda) +\log{ {1\over 2} +
\left ( {1\over 2} - \lambda \right )
{\rm erf}\left ( {{1\over 2} - \lambda \over\sqrt{2\gamma}} \right )}
\label{g1lambda}
\ee
We show the shape of the the function $G_1(\lambda)$ in Fig. \ref{g1}.
Evaluating the integral (\ref{number1}) with the saddle-point
method one gets 
\be
n(N)\simeq  C\, e^{0.14 N}.
\label{nestimate}
\ee
Notice that the exponent is the half 
of what is observed numerically (cfr (\ref{ntrue})). In the remaining part of the paper
we shall argue that (\ref{nestimate}) is indeed an estimate from below of the
actual number of metastable states.
\vskip 0.2cm
\noindent
The above discussion has been able to reproduce the
exponential growth of the number of metastable states 
with the size of the system. 
To understand the discrepacy between the estimated
exponent and the one measured numerically 
one should note that the nature of the interaction makes the 
conditional probabilities play a major role in the asymptotics
of the number of metastable states. 
To be more precise,
our approximation which assumes mutually independent individual $\o$-stability events, i.e.
$\pi_\lambda(\omega_{k+1}\vert\omega_1,\ldots,\omega_k)\simeq
\pi_{\lambda}(\omega_{k+1})$, is actually reasonable only for small
value of $k$ (this can be checked, for example, calculating
the correlation functions). As numerical results shows, 
for large values of $k$ the specific form 
of the interactions make these events strongly dependent.
In Fig. \ref{conditional} we show the function $P(k)$ providing 
the average of 
$\pi_\lambda(\omega_{k+1}\vert\omega_{1},\ldots,\omega_k)$
over a large sample of different
permutations
$(\omega_1,\ldots,\omega_N)$ of  the lattice points. The conditional probabilities
$P(k)$ grow monotonically, almost linearly, from the initial 
(unconditioned) value up to a number close to $1$.
In other words, requiring that a large number $k$ of 
spins produce an $\o$-stable state increases substantially the probability of doing the same
for the remaining spins.
\vskip 0.2cm

\begin{figure}[ht]
\includegraphics[width=5.5in, angle=-90]{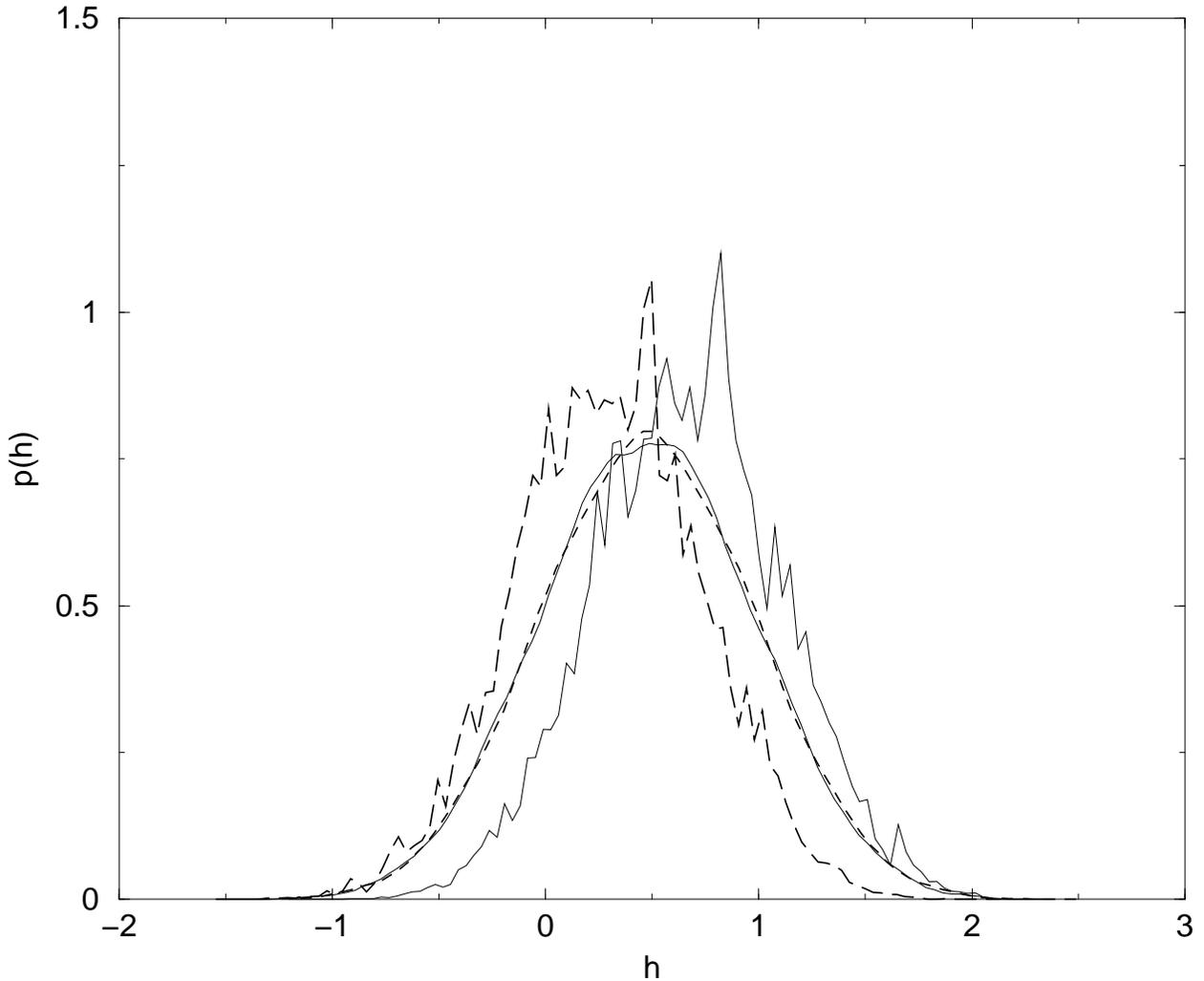}
\caption{\baselineskip=13pt {\small Graphs of the four distributions of
the values of $h$ described in the text.}}
\protect\label{acca}
\end{figure}

\noindent
Another way of understanding the constructive effect of the 
correlations 
is the following. Consider again the function $h(\tau,\omega)$. Having fixed
$\omega_{k+1}$, we have already noticed that for $s$ and $N$ large enough the values of
$h(\tau,\omega_{k+1})$ with
$\tau\in\Omega_s$ are approximately distributed according to a gaussian probability density
with mean $\lambda =s/N$ and variance $\gamma =\lambda(1-\lambda)$, regardless of the
particular value of $\omega_{k+1}$. Thus, in particular, the same distribution 
are expected to arise if one considers the values of
$h(\tau,\omega_{k+1})$ constrained to the subsets of configurations such that
$\omega_{k+1}\in\tau$, or 
$\omega_{k+1}\notin\tau$.
On the other hand, if one picks $\omega_1,\omega_2,\ldots,\omega_{k}$ with 
$\omega_j\neq\omega_{k+1}$, $j=1,\ldots,k$, and computes numerically 
the two conditional distributions of the values of $h(\tau,\omega_{k+1})$
given $\omega_1$-stability, ... , $\omega_k$-stablity (again with the constraints
$\omega_{k+1}\in\tau$ or 
$\omega_{k+1}\notin\tau$), one finds that their means move to opposite directions, 
thus increasing the probability of $\omega_{k+1}$-stability (see (\ref{totalprob})).
This is  shown in Fig. \ref{acca}, where a system of size $N=21$ and $s = 10$
is considered.
The two central distributions 
correspond to the unconditioned cases, namely the values of 
$h(\tau, \omega)$ for $\tau \in \Omega_{10}$ with the only constraints 
$\omega\in\tau$ or $\omega\notin\tau$, respectively. Considering
instead the values taken by $h$ on the states $\tau$ which, besides 
the constraints specified above, are  stable with respect the first 
$10$ spins, one finds two distributions whose mean values have moved 
towards opposite directions. An averaged over $\omega$ has been performed. 
\section{Conclusion}
We have investigated the statistical properties
of energy levels and metastables states for a class of deterministic
models, the most representative being the {\em sine} model\cite{MPR}, 
which  have attracted much attention in recent years for
their glassy behaviour despite the non-random nature of the 
interaction. We have pushed further on the analogy with glassy systems,
proving a number of properties typical of disordered spin models. 
In particular, using number theoretic methods, we have 
described the energy (equivalently, free energy at $T=0$) landscape 
as a function of configurations with a fixed overlap with the ground 
state. The analysis revealed the existence of states very different 
from the ground state but with energy arbitrarily close to it: this
corresponds to the ``chaoticity'' property of spin-glasses systems,
well established in long range models. More importantly, some of these
states can be local energy minima (equivalently, $1$-flip stable at $T=0$).
They
 are expected to have a significant weight on the partition function
in the low-temperature region, giving rise to the 
non-equibrium behaviour observed in annealing Montecarlo experiments.
We have been able to estimate the approximate number of
these energy minima. The analytic computations, combined with the
numerical findings, strongly support the conclusion that the
 bound (\ref{nestimate})  estimates from below 
the number of metastable states $n(N)$,
 proving their exponential increase
with the size of the system.

A number of basic questions about
metastability
 arises now in a natural way, such as computing the
energy density 
distributions of metastables states, 
studying energy barriers among them and their attraction basins.
Stability of configurations with respect to the flip of an arbitrary
 number
of spins is an interesting question as well. These problems
 are currently
under investigation using the approach developed in this 
 paper and will be
addressed elsewhere.

\section*{Appendix}
\begin{itemize}

\item{(Proof of (\ref{gs}))}
Choose $N$ odd such that $p = 2N+1$ is prime of the form $4m+3$, where 
$m$ is
an integer. Denote by $\s^0$ the spin configuration  given by the 
sequence of Legendre symbols, i.e.
$$ \s^{0}_x = \left (\frac{x}{p}\right ) = \left\{ \begin{array}{ll}
                                            +1, & \mbox{if $x=k^2 
(\bmod p)$}, \\
                                            -1, & \mbox{if $x \neq k^2 
(\bmod p)$},        
                                             \end{array}
                                     \right.$$
with $k=1,2,\ldots,p-1$. 
Let us show that $\s^0=(\s_1^{0}\ldots\s_N^{0})$ is the ground
state for the sine model or, which is the same,
that $\s^0$ is an eigenvector of $J$ with  eigenvalue $1$. For basic results of
number theory used in the proof see, for example, ref. \cite{Ap}.
\begin{eqnarray*}
(J\s^0)_y & = & \frac{2}{\sqrt{p}}\sum_{x=1}^N 
              \sin \left( \frac{2\pi xy}{p}\right)\left 
              (\frac{x}{p}\right ) \\
        & = & \frac{2}{\sqrt{p}}\sum_{x=1}^N 
              \frac{1}{2i}\left 
              (\exp \left( \frac{i2\pi xy}{p}\right)\left 
(\frac{x}{p}\right )
            -  \exp \left( \frac{-i2\pi xy}{p}\right)\left 
(\frac{x}{p}\right )
              \right ) \\
        &   & \\
        &   & \mbox{changing $x\mapsto -x$ in the second summation}\\
        &   & \\
        & = & \frac{1}{i \sqrt{p}}\left [ \sum_{x=1}^N 
              \exp \left( \frac{i2\pi xy}{p}\right)\left 
(\frac{x}{p}\right )
              - \sum_{x=-N}^{-1}
              \exp \left( \frac{i2\pi xy}{p}\right)\left 
(\frac{-x}{p}\right )
              \right ] \\
        &   & \\
        &   & \mbox{using multiplicativity of Legendre symbols:}
            \left (\frac{-x}{p}\right ) = \left (\frac{-1}{p}\right 
)\left (\frac{x}{p}\right )\\
             &  &  \mbox{and the fact that $\left (\frac{-1}{p}\right ) 
= -1$ if
              $p=3 (\bmod 4)$ }\\
        &   & \\
        & = & \frac{1}{i \sqrt{p}}\left [ \sum_{x=1}^N 
              \exp \left( \frac{i2\pi xy}{p}\right)\left 
(\frac{x}{p}\right )
              + \sum_{x=-N}^{-1}
              \exp \left( \frac{i2\pi xy}{p}\right)\left 
(\frac{x}{p}\right )
              \right ] \\
        &   & \\      
        &   & \mbox{using the periodicity of Legendre' symbols}: \,
              \left (\frac{x+p}{p}\right ) = \left (\frac{x}{p}\right ) 
\\
        &   & \\       
        & = & \frac{1}{i \sqrt{p}}\left [ \sum_{x=1}^N 
              \exp \left( \frac{i2\pi xy}{p}\right)\left 
(\frac{x}{p}\right )
            + \sum_{x=N+1}^{2N}
            \exp \left( \frac{i2\pi xy}{p}\right)\left 
(\frac{x}{p}\right )
              \right ] \\
        &   & \\      
        &   & \mbox{being $\left (\frac{p}{p}\right ) = 0$ by 
definition} \\
        &   & \\       
        & = & \frac{1}{i \sqrt{p}}\sum_{x=1}^{p} 
              \exp \left( \frac{i 2\pi xy}{p}\right) \left 
              (\frac{x}{p}\right ) \\
        &   & \\      
        &   & \mbox{using the separability for Gauss sums } \\
        &   & \\           
        & = & \frac{1}{i \sqrt{p}}\left( \frac{y}{p}\right )
             \sum_{x=1}^{p} 
             \exp \left( \frac{i 2\pi x}{p}\right) \left
             (\frac{x}{p}\right ) \\
        &   & \\      
        &   & \mbox{evaluating the Gauss sum } \\
        &   & \\     
        & = & \frac{1}{i \sqrt{p}}\left( \frac{y}{p}\right )i \sqrt{p}   
\; = \; \s^0_y\,,
      \qquad \qquad \qquad  \qquad
\end{eqnarray*}
which is the desired property.
\vskip 0.5 cm
\noindent

\item{(Proof of (\ref{moment1}) and (\ref{moment2}))}
We sketch the basic steps
of the calculation. Set
$$
\alpha= -\frac{N}{2} +2s ~~~~~~~~~~~~~~
c_j = \binomial{N-j}{s-j}.
$$
We then have 
$$
\E_s(H^n) = {1\over c_0}\sum_{\tau\in\W} \left( \alpha 
-2\sum_{x\in\tau}
              \sum_{y\in\tau}J_{xy} \right )^n 
         =  {1\over c_0}\sum_{\tau\in\W}\sum_{k=0}^n\binomial{n}{k}
              \alpha^{n-k} \left (- 
2\sum_{x\in\tau}\sum_{y\in\tau}J_{xy}
              \right )^k.
$$
\noindent 
For $n=1$ we get
\begin{eqnarray*}
\sum_{\tau\in\W}\left (\sum_{x\in\w}\sum_{y\in\w}J_{xy}\right )   
    & = & \sum_{x=1}^N\sum_{y\neq x,y=1}^N c_2 J_{xy}\s_x\s_y
          +\sum_{x=1}^N c_1 J_{xx} \\
    & = & \sum_{x=1}^N\sum_{y=1}^N c_2 J_{xy}\s_x\s_y
          +\sum_{x=1}^N (c_1-c_2) J_{xx} = c_2(N-1) +c_1,    
\end{eqnarray*}
whereas for $n=2$ we have
\begin{eqnarray*}
&&\sum_{\w\in\W}\left (\sum_{x\in\w}\sum_{y\in\w}J_{xy}\right )^2 
     =     \sum_{z=1}^N\sum_{u\neq z;u=1}^N J_{zu}\s_z\s_u
           \left [
            \sum_{x\neq z,u; x=1}^N\sum_{y\neq z,u,x; y=1}^N 
c_4J_{xy}\s_x\s_y
           \right.\\
  &+&\sum_{y\neq z,u; y=1}^N c_3J_{zy}\s_z\s_y
            +\sum_{y\neq z,u; y=1}^N c_3J_{uy}\s_u\s_y 
    +\sum_{x\neq z,u; x=1}^N c_3J_{xz}\s_x\s_z
            +\sum_{x\neq z,u; x=1}^N c_3J_{xu}\s_x\s_u \\
            &+& \left.\sum_{x\neq z,u; x=1}^N c_3J_{xx}
            +c_2J_{zu}\s_z\s_u
            +c_2J_{zz}
            +c_2J_{uu}
           \right ] + \sum_{z=1}^NJ_{zz} 
           \left [ 
            \sum_{x\neq z; x=1}^N\sum_{y\neq z,x; y=1}^N 
c_3J_{xy}\s_x\s_y \right.\\
            &+&\sum_{x\neq z; x=1}^N c_2J_{xz}\s_x\s_z
            \left.
            +\sum_{y\neq z; y=1}^N c_2J_{yz}\s_y\s_z 
            +\sum_{x\neq z; x=1}^N c_2J_{xx}
            +c_1J_{zz}
           \right ]\\ 
    & = &  c_4(N-1)(N-3)+2c_3(N-1)+2c_2(N-1)+c_1,
\end{eqnarray*}
which easily give the desired identities.
\item (Proof of (\ref{variancemagn})) Let us first extend everything to the
set $\{1,2,\ldots,p-1\}$ which, $p$ being prime, is a number field. Here we can
exploit the multiplicative structure of the field and of the `extended ground
state'
$\sigma^0_x=\legendre{x}{p}$,
$x=1,\ldots, q$, with $q=p-1$.
With slight abuse of notation we shall use the same symbols $\O_s$, $p_s$ and
$\E_s$ to denote the corresponding extended quantities.
It is immediate to see that $m(\s^0)=0$ and
$$
E_s(m)\,=\,\frac{1}{\binomial{q}{s}}\sum_{\tau\in\Omega_s}
\left(\frac{-2}{q}\sum_{x\in\tau}\legendre{x}{p}\right)=0.
$$
In order to calculate the second moment we consider 
\begin{eqnarray*}
\frac{1}{\binomial{q}{s}}\sum_{\tau\in\Omega_s}\left(\sum_{x\in\tau}\legendre{x}{p}\right)^2\,&=&\,\frac{1}{\binomial{q}{s}}\sum_{\tau\in\Omega_s}\left(\sum_{x\in\tau^2}\legendre{x}{p}\right)\\
&=&\frac{1}{\binomial{q}{s}}\sum_{x=1}^q c_p(x)\,\legendre{x}{p}\, ,
\end{eqnarray*}
where, for any given $\tau\in\Omega_s$, $\tau^2$ is the collection, with multiplicity, of all possible products $x_j\cdot x_i$, $x_i, x_j\in\tau$ (all the operations are $\mbox{mod } p$). For example, if $\tau=\{x_1,x_2,x_3\}$ then
$\tau^2=\{x_1^2,x_2^2,x_3^2, x_1x_2,x_1 x_3,x_2x_1,x_3 x_1,x_2 x_3,x_3 x_2\}$.
Also, for any given $x\in\{1,\ldots,q\}$,
$$
c_p(x)\,=\,\sum_{\tau\in\Omega_s}\{\mbox{number of times } x\in\tau^2\}=
\sum_{u=1}^{q}\sharp\{\tau\,\vert\,u\in\tau \mbox{ and
}u^{-1}x\in\tau\}.
$$
In particular, if $\legendre{x}{p}=-1$ then $u\neq u^{-1}x$, $\forall
u=1,\ldots,q$, therefore
$$
c_p(x)\,=\,\sum_{u=1}^{q} \binomial{q-2}{s-2}\,=\, q\binomial{q-2}{s-2}.
$$
If instead $\legendre{x}{p}=1$, then there exists ${\bar {u}}$ such that
 ${\bar{u}}^2=x$, i.e.
$$
c_p(x)\, =\, \sum_{u\neq\pm{\bar{u}}}\binomial{q-2}{s-2} +
2\binomial{q-1}{s-1}\,=\,(q-2)\binomial{q-2}{s-2} +2\binomial{q-1}{s-1}.
$$
Putting everything together we get the following expression for the variance
$\sigma^2_s(m)$:
\begin{eqnarray*}
\sigma^2_s(m)&=&
\frac{4}{q^2\binomial{q}{s}}\sum_{\tau\in\Omega_s}\left(\sum_{x\in\tau}\legendre{x}{p}\right)^2\\
&=&\frac{4}{q^2\binomial{q}{s}}\left[-\frac{q^2}{2}\binomial{q-2}{s-2}+\frac{q}{2}(q-2)\binomial{q-2}{s-2}
+ q\binomial{q-1}{s-1}\right]\\
&=&\frac{4}{q\binomial{q}{s}}\left[\binomial{q-1}{s-1}-\binomial{q-2}{s-2}\right]={4s(q-s)
\over q^2(q-1)}.
\end{eqnarray*}
We now turn back to our the original lattice $\{1,\dots, N\}$.
Again we can write
$$
\frac{1}{\binomial{N}{s}}\sum_{\tau\in\Omega_s}\left(\sum_{x\in\tau}\legendre{x}{p}\right)^2\,=
\,\frac{1}{\binomial{N}{s}}\sum_{x=1}^{p} c_N(x)\legendre{x}{p}.
$$
In this case, however, the multiplicity function $c_N(x)$ can not be
handled as easily as before. In particular, given $x\in\{1,\ldots N\}$, we denote by
$\Gamma(x)$ the set
$=\{u_1,\ldots,u_{d_N(x)}\}$ given by the $u$'s in $\{1,\ldots,N\}$ such
that $u^{-1}x\in\{1,\ldots,N\}$. The cardinality $d_N(x)$ of the set $\Gamma(x)$ is a
non trivial function of $x$. It is  shown in Fig. \ref{dN} for $1\leq x \leq N$.
If now $\legendre{x}{p}=-1$, then clearly
$$
c_N(x)\,=\,d_N(x)\cdot\binomial{N-2}{s-2}.
$$
On the other hand, if $\legendre{x}{p}=1$ (i.e. ${\bar u}^2=x$), then 
(note that either ${\bar u}\in\{1,\ldots,N\}$ or
$-{\bar u}\in\{1,\ldots,N\}$)
$$
c_N(x)\,=\, (d_N(x)-1)\cdot\binomial{N-2}{s-2}+\binomial{N-1}{s-1} .
$$

\begin{figure}[ht]
\includegraphics[width=2.7in]{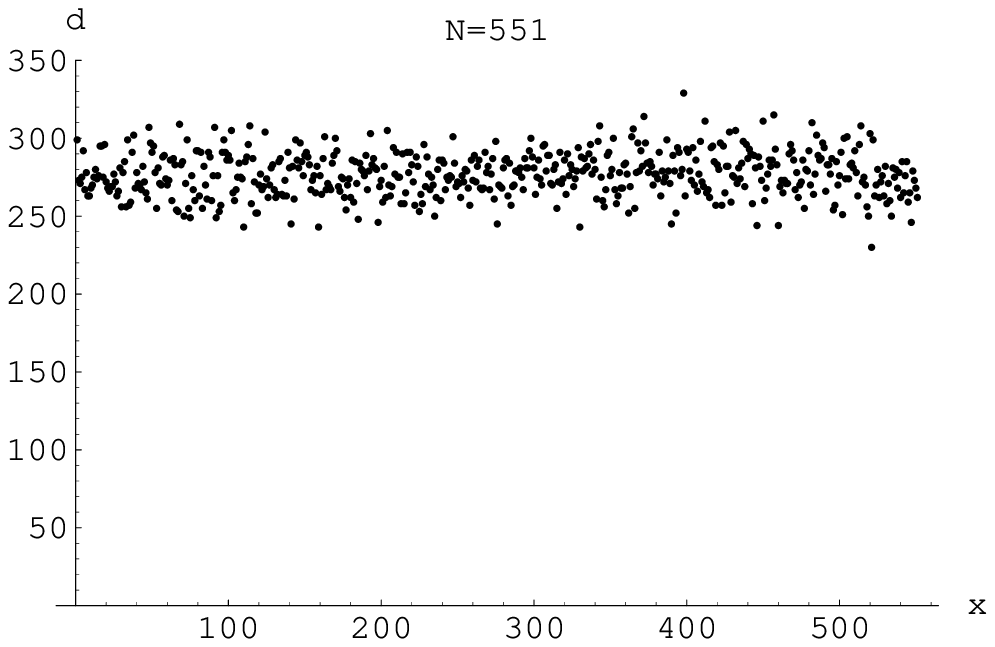}
\includegraphics[width=2.7in]{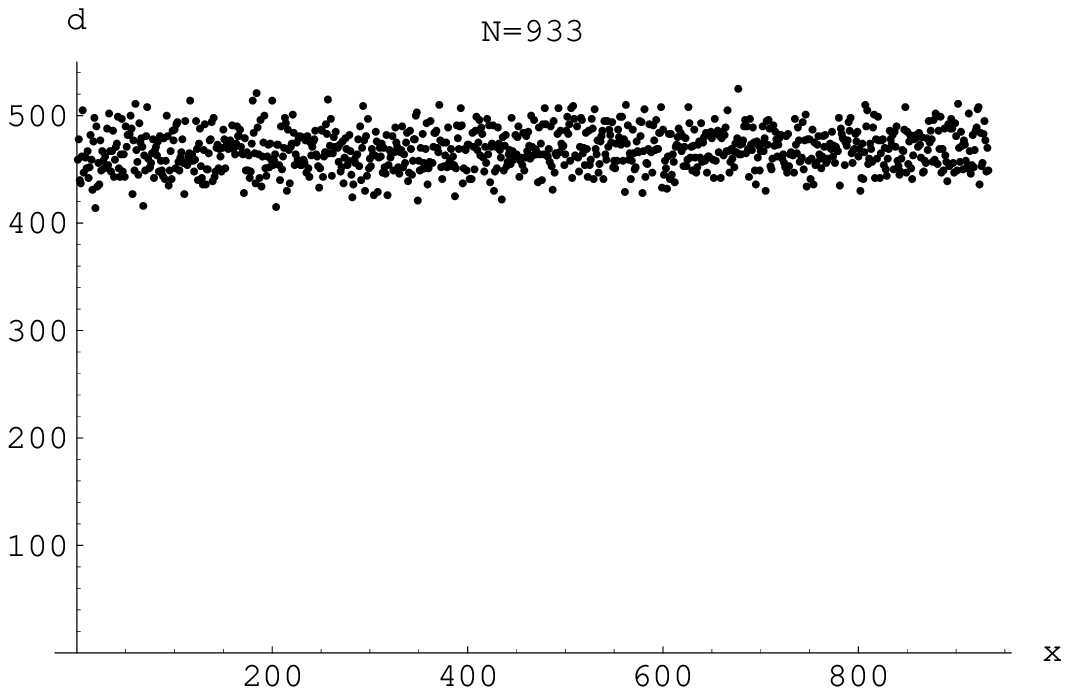}
\caption{\baselineskip=13pt {\small {The function $d_N(x)$ versus $x$, for $N=551$ and 
$N=933$}}}
\protect\label{dN}
\end{figure}
We can then use these informations and write

\begin{eqnarray*}
\frac{1}{\binomial{N}{s}}\sum_{\tau\in\Omega_s}\left(\sum_{x\in\tau}\legendre{x}{p}\right)^2&=&
\frac{1}{\binomial{N}{s}}\sum_{x=1}^{p} c_N(x)\,\legendre{x}{N}\\
&=& \frac{1}{\binomial{N}{s}}\left[-\sum_{\legendre{x}{p}=-1} d_N(x)\cdot\binomial{N-2}{s-2}\,+\right.\\
&&\left.+\,\sum_{\legendre{x}{p}=1}
\left(\,(d_N(x)-1)\cdot\binomial{N-2}{s-2}+\binomial{N-1}{s-1}\,\right)\right]\\
&=&\frac{1}{\binomial{N}{s}}\left[\binomial{N-2}{s-2}\cdot\sum_{x=1}^p
d_N(x)\legendre{x}{p}+\right.\\ &&\left.+
\left( \binomial{N-1}{s-1}-\binomial{N-2}{s-2}\right)\cdot\sum_{\legendre{x}{p}=1}1\right].
\end{eqnarray*}
\vskip 0.3cm
\noindent
Finally, we have the following expression for the variance $\sigma_s^2(m)$  of the
magnetization $m$ over the space $\Omega_s$:
$$
\sigma_s^2(m)=\frac{4}{N^2}\,
\E_s\left(\biggl(\sum_{x\in\tau}\legendre{x}{p}\biggr)^2\right)\, -\, {4s^2(m^0)^2\over
N^2},
$$
from which one easily gets formula (\ref{variancemagn}).

\end{itemize}

\end{document}